\documentstyle[12pt]{article}

\def\fund{  \> {\vcenter  {\vbox
               {\hrule height.6pt
                \hbox {\vrule width.6pt  height5pt
                      \kern5pt
                      \vrule width.6pt  height5pt}
                \hrule height.6pt}
                         }
               }
            \>\>  }

\def\antifund{  \> \overline{ {\vcenter  {\vbox
               {\hrule height.6pt
                \hbox {\vrule width.6pt  height5pt
                      \kern5pt
                      \vrule width.6pt  height5pt}
                \hrule height.6pt}
                         }
               } }
            \>\>  }

\def\sym{  \> {\vcenter  {\vbox
              {\hrule height.6pt
               \hbox {\vrule width.6pt  height5pt
                      \kern5pt
                      \vrule width.6pt  height5pt
                      \kern5pt
                      \vrule width.6pt  height5pt}
               \hrule height.6pt}
                         }
               }
            \>\>  }

\def\symbar{  \> \overline{ {\vcenter  {\vbox
              {\hrule height.6pt
               \hbox {\vrule width.6pt  height5pt
                      \kern5pt
                      \vrule width.6pt  height5pt
                      \kern5pt
                      \vrule width.6pt  height5pt}
               \hrule height.6pt}
                         }
               } }
            \>\>  }

\def\antisym{ \> {\vcenter  {\vbox
                 {\hrule height.6pt
                  \hbox {\vrule width.6pt  height5pt
                         \kern5pt
                         \vrule width.6pt  height5pt}
                  \hrule height.6pt
                  \hbox {\vrule width.6pt  height5pt
                         \kern5pt
                         \vrule width.6pt  height5pt}
               \hrule height.6pt}
                         }
               }
            \>\>  }

\def\antithree{ \> {\vcenter  {\vbox
                 {\hrule height.6pt
                  \hbox {\vrule width.6pt  height5pt
                         \kern5pt
                         \vrule width.6pt  height5pt}
                  \hrule height.6pt
                  \hbox {\vrule width.6pt  height5pt
                         \kern5pt
                         \vrule width.6pt  height5pt}
                  \hrule height.6pt
                  \hbox {\vrule width.6pt  height5pt
                         \kern5pt
                         \vrule width.6pt  height5pt}
               \hrule height.6pt}
                         }
               }
            \>\>  }

\def\antifour{ \> {\vcenter  {\vbox
                 {\hrule height.6pt
                  \hbox {\vrule width.6pt  height5pt
                         \kern5pt
                         \vrule width.6pt  height5pt}
                  \hrule height.6pt
                  \hbox {\vrule width.6pt  height5pt
                         \kern5pt
                         \vrule width.6pt  height5pt}
                  \hrule height.6pt
                  \hbox {\vrule width.6pt  height5pt
                         \kern5pt
                         \vrule width.6pt  height5pt}
                  \hrule height.6pt
                  \hbox {\vrule width.6pt  height5pt
                         \kern5pt
                         \vrule width.6pt  height5pt}
               \hrule height.6pt}
                         }
               }
            \>\>  }

\def\antifive{ \> {\vcenter  {\vbox
                 {\hrule height.6pt
                  \hbox {\vrule width.6pt  height5pt
                         \kern5pt
                         \vrule width.6pt  height5pt}
                  \hrule height.6pt
                  \hbox {\vrule width.6pt  height5pt
                         \kern5pt
                         \vrule width.6pt  height5pt}
                  \hrule height.6pt
                  \hbox {\vrule width.6pt  height5pt
                         \kern5pt
                         \vrule width.6pt  height5pt}
                  \hrule height.6pt
                  \hbox {\vrule width.6pt  height5pt
                         \kern5pt
                         \vrule width.6pt  height5pt}
                  \hrule height.6pt
                  \hbox {\vrule width.6pt  height5pt
                         \kern5pt
                         \vrule width.6pt  height5pt}
               \hrule height.6pt}
                         }
               }
            \>\>  }

\def\antisix{ \> {\vcenter  {\vbox
                 {\hrule height.6pt
                  \hbox {\vrule width.6pt  height5pt
                         \kern5pt
                         \vrule width.6pt  height5pt}
                  \hrule height.6pt
                  \hbox {\vrule width.6pt  height5pt
                         \kern5pt
                         \vrule width.6pt  height5pt}
                  \hrule height.6pt
                  \hbox {\vrule width.6pt  height5pt
                         \kern5pt
                         \vrule width.6pt  height5pt}
                  \hrule height.6pt
                  \hbox {\vrule width.6pt  height5pt
                         \kern5pt
                         \vrule width.6pt  height5pt}
                  \hrule height.6pt
                  \hbox {\vrule width.6pt  height5pt
                         \kern5pt
                         \vrule width.6pt  height5pt}
                  \hrule height.6pt
                  \hbox {\vrule width.6pt  height5pt
                         \kern5pt
                         \vrule width.6pt  height5pt}
               \hrule height.6pt}
                         }
               }
            \>\>  }

\begin{document}
\setlength{\baselineskip}{18pt}

\vspace*{5mm}

\begin{flushright}
DPNU-98-22 \\
hep-ph/9806273 \\
June 1998 
\end{flushright}

\vspace{1cm}

\begin{center}
{\large\bf A New Model of Direct Gauge Mediation 
with an Affine Quantum Moduli Space}
\end{center}

\vspace{1cm}

\begin{center}
{\bf Nobuhito Maru}\footnote{Research Fellow of 
the Japan Society for the Promotion of Science} 
\end{center}

\begin{center}
{\em Department of Physics, 
Nagoya University \\ 
Nagoya 464-8602, JAPAN} 
\end{center}

\begin{center} 
{\normalsize {\tt maru@eken.phys.nagoya-u.ac.jp}} 
\end{center}

\vspace{1cm}
\setcounter{page}{0}
\thispagestyle{empty}
\begin{center}
{\bf Abstract}
\end{center}
We present a new, simple supersymmetry 
breaking model 
with direct gauge mediation. 
Supersymmetry breaking sector is based on the 
model with an Affine quantum moduli space. 
The model has no gauge messengers and 
no light scalars charged under 
the Standard Model which give a negative 
contribution to the soft masses. 
Large expectation value at the minimum is 
obtained by balancing the runaway potential 
and the dimension six nonrenormalizable term. 
This makes it easy to preserve 
the perturbative unification. 
Enough suppression of 
the supergravity contribution 
to the soft masses can also be shown.   

\vspace{1cm}

\begin{flushleft}
PACS: 11.30.Pb, 12.60.Jv \\
Keywords: Dynamical Supersymmety Breaking, 
Direct Gauge Mediation, \\
\hspace{60pt}Affine Quantum Moduli Space
\end{flushleft}

\newpage

Gauge mediated supersymmetry (SUSY) breaking 
\cite{DN} has been much considered 
as an attractive alternative 
to the supergravity (SUGRA) mediation 
since the degeneracy of the sfermions is 
automatically guaranteed and the soft parameters 
can be described in terms of a few parameters. 
However, the original model \cite{DN} is quite 
compicated, especially there is 
an unnatural sector 
called ``messenger sector". 
There has been much progress to 
simplify the structure of the model 
along various lines 
\cite{HIY}-\cite{CRS}. 
The simplest idea is ``Direct Gauge 
Mediation"(DGM). 
Recently, many authors have proposed 
interesting DGM models 
\cite{PT}-\cite{Agashe}. 
For the reader who would like to know about 
these developments in more detail, 
we recommend the beautiful review of 
Giudice and Rattazzi \cite{GR}.

In this article, we present a new, 
simple SUSY breaking model 
with direct gauge mediation.
\footnote{Our model has similar dynamics 
to the model in Ref. \cite{Shirman}} 
The symmetries of our model are 
\begin{equation}
SU(6)_1 \times SU(6)_2 \times [SU(6)], 
\end{equation}
where the first two $SU(6)_{1,2}$ are gauge groups 
and $SU(6)$ in the bracket is a global symmetry. 
We will later identify the subgroup of this 
$SU(6)$ with the Standard Model (SM) 
gauge groups.

The field content of our model is as follows. 
\begin{eqnarray}
\label{matter}
X &\sim& (\antithree, {\bf 1}; {\bf 1}), 
\nonumber \\
\Sigma &\sim& (\fund, \antifund; {\bf 1}), 
\nonumber \\
Q &\sim& (\antifund, {\bf 1}; \fund), 
\nonumber \\
\bar{Q} &\sim& ({\bf 1}, \fund; \antifund). 
\end{eqnarray}
We note that this field content consists of 
two sectors. 
The first sector includes $X$ only. 
This model is $SU(6) + \antithree$, 
which has been originally discussed 
by Cs\'aki, Schmaltz and Skiba 
\cite{CSS} and also discussed recently 
in Ref. \cite{CM,DM}. 
According to Ref. \cite{CSS}, 
the low energy effective theory of this model 
has two branches, 
one is $W_{dyn} = 0$ 
(``an Affine quantum moduli space") 
and another is $W_{dyn} \ne 0$ 
(due to gaugino condensation). 
The definition of Affine quantum moduli space is 
that the moduli space is given by gauge invariant 
polynomials with no relations among them 
and $W_{dyn}=0$. 
Moreover, 't Hooft anomalies between fundamental 
fields and gauge invariant composites match 
at all points in the moduli space 
(including the origin). 
Therefore, if the appropriate gauge invariant 
operators to lift the flat directions 
are added to 
the superpotential, then we can expect the model 
to break SUSY due to confinement such as the model 
proposed by Intriligator, Seiberg and Shenker 
(ISS) \cite{ISS}. 
We refer to this sector as SUSY breaking sector 
throughout this paper. 
We will utilize this branch to construct the model 
of direct gauge mediation. 
On the other hand, the case of $W_{dyn} \ne 0$, 
we will discuss later.

The second sector contains $\Sigma, Q, \bar{Q}$. 
These superfields are necessary to communicate 
SUSY breaking effects to the observable sector. 
The field content for each gauge group is 
$SU(6)_{1,2} + 6(\fund + \antifund)$, 
which is the special case of the model 
in Ref. \cite{PST}.

Note also that this model is completely chiral, 
in other words, 
we cannot add mass terms for any field 
to the superpotential.

%\begin{table}[hbtp]
%\begin{center}
%\begin{tabular}{|c|cc|c|}
%\hline
%& $SU(6)_1$ & $SU(6)_2$ & $SU(6)$ \\ 
%\hline
%\vphantom{\Big(}
%$X$ & $\antithree$ & ${\bf 1}$ & ${\bf 1}$ \\ 
%\hline
%$\Sigma$ & $\fund$ & $\antifund$ & ${\bf 1}$ \\ 
%$Q$ & $\antifund$ & ${\bf 1}$ & $\fund$ \\ 
%$\bar{Q}$ & ${\bf 1}$ & $\fund$ & $\antifund$ \\ 
%\hline
%\end{tabular}
%\caption[matter]{Matter content of the model}
%\end{center}
%\end{table}

Here we take the following tree level 
superpotential
%\footnote{Explicit contraction 
%of indices in $X^4$ are $X^4 \equiv 
%X_{a_{1}b_{1}c_{1}}X_{d_{1}e_{1}f_{1}}
%X_{a_{2}b_{2}c_{2}}X_{d_{2}e_{2}f_{2}}
%\epsilon^{a_{1}b_{1}f_{1}a_{2}b_{2}f_{2}}
%\epsilon{d_{1}e_{1}c_{1}d_{2}e_{2}c_{2}}$} 
%
\begin{equation}
\label{tree}
W = \lambda_1 \Sigma Q \bar{Q} + 
\frac{\lambda_2}{M_P} X^4 + 
\frac{\lambda_3}{M_P^3} {\rm det}\Sigma, 
\end{equation}
where $\lambda_{1,2,3}$ are the couplings of 
order unity and $M_P$ is the reduced Planck scale. 
Although it is possible to add other 
nonrenormalizable terms to the superpotential, 
we forbid them by imposing symmetries.

In the presence of the first two terms 
in the superpotential, 
there exist some classical flat directions: 
$v^6 \equiv {\rm det}\Sigma, B \equiv Q^6, 
B \equiv \bar{Q}^6, M \equiv XQ^3$.

Let us first discuss the classical direction 
$\langle \Sigma \rangle \ne 0$ 
we are interested in, 
which corresponds to the direction 
${\rm det}\Sigma \ne 0$;
\begin{equation}
\label{vev}
\langle \Sigma \rangle = 
{\rm diag} (v,v,v,v,v,v). 
\end{equation}
Along this direction, the gauge group 
$SU(6)_1 \times SU(6)_2$ is broken to 
their diagonal $SU(6)_D$ 
and $Q, \bar{Q}$ become massive 
since the superpotential includes 
the mass term 
$\lambda_1 \langle \Sigma \rangle Q \bar{Q}$. 
For large $v$, 
the low energy effective theory is 
$SU(6)_D + \antithree +$ one singlet $v$. 
\footnote{We use the same notation 
$\langle \Sigma \rangle$ 
for a singlet superfield.} 
As mentioned earlier,
the low energy dynamics of this model has 
already been discussed \cite{CSS}. 
$U(1)_R$ anomaly of the fermionic component of $X$, 
gaugino and that of the fermionic component of 
$X^4$ are saturated. 
This implies that the theory is in the confining 
phase below the scale $\Lambda_L$ 
(the strong coupling scale of $SU(6) + \antithree$) 
and the low energy effective theory should be 
described in terms of the composite $X^4$. 
The form of the dynamically generated superpotential 
due to gaugino condensation from 
the subgroup $SU(3) \times SU(3)$ in $SU(6)$ is 
\begin{equation}
\label{33}
W_{dyn} = (\omega^r-\omega^s) \Lambda_3^3,
\end{equation}
where $\omega$ represents the cube root of unity 
and $\Lambda_3$ means the strong coupling scale of 
pure $SU(3)$. 
For details of the derivation of (\ref{33}), 
we refer the reader to Ref. \cite{DM}.

What is important here is that this model 
has $W_{dyn}=0$ branch. 
In this situation, the effective superpotential 
becomes 
\begin{equation}
\label{eff}
W_{eff} = \frac{\lambda_2}{M_P}u, \quad u \equiv X^4. 
\end{equation}
Taking into account that the effective K\"ahler 
potential is 
\begin{equation}
K_{eff} \sim u^{\dag} u/|\Lambda_L|^6, 
\end{equation}
we obtain the following scalar potential. 
\begin{equation}
\label{scapot}
V_{eff} = \left( \frac{\partial^2 K_{eff}}
{\partial u^{\dag} \partial u} \right)^{-1} 
\left| \frac{\partial W_{eff}}{\partial u} 
\right|^2 = \frac{\lambda^2_2}{M_P^2} 
\Lambda_L^6. 
\end{equation}
At the first glance, one may think that this model 
is the plateau model \cite{Murayama,DDRG,Agashe} 
since the scalar potential is flat. 
However, we have to recall that $\Lambda_L$ 
depends on the singlet superfield $v$ 
through 1-loop matching of 
the gauge couplings: 
\begin{equation}
\left( \frac{\Lambda_1}{v} \right)^9 
\left( \frac{\Lambda_2}{v} \right)^{12} = 
\left( \frac{\Lambda_L}{\lambda_1 v} \right)^{15}, 
\end{equation}
where $\Lambda_{1,2}$ denotes the dynamical scale of 
$SU(6)_1 + \antithree + 6 (\fund + \antifund)$ and 
$SU(6)_2 + 6 (\fund + \antifund)$, respectively. 
Then, the scalar potential (\ref{scapot}) becomes 
\begin{equation}
V_{eff} = \frac{\lambda_2^2}{M_P^2} 
\left( \frac{\lambda_1^5 \Lambda_1^3 \Lambda_2^4}
{v^2} \right)^{\frac{6}{5}} = 
\frac{\lambda_2^2}{M_P^2} 
\left( \frac{\lambda_1^5 \Lambda^7}{v^2} 
\right)^{\frac{6}{5}}, 
\end{equation}
where $\Lambda$ is defined as $\Lambda^7 \equiv 
\Lambda_1^3 \Lambda_2^4$ for simplicity. 
This results in runaway behavior but 
the term ${\rm det} \Sigma$ in Eq. (\ref{tree}) 
stabilizes the scalar potential.

Indeed, by minimizing the scalar potential 
\begin{equation}
V_{eff} = \frac{\lambda_2^2}{M_P^2} 
\left( \frac{\lambda_1^5 \Lambda^7}
{v^2} \right)^{\frac{6}{5}}  + 
\frac{36\lambda_3^2}{M_P^6} v^{10}, 
\end{equation}
we obtain 
\begin{equation}
v \sim (\Lambda^{21} M_P^{10})^{\frac{1}{31}}, \quad 
F_v \sim \left(\frac{\Lambda^{105}}{M_P^{43}} 
\right)^{\frac{1}{31}}, \quad 
V_0 \sim \left(\frac{\Lambda^{210}}{M_P^{86}} 
\right)^{\frac{1}{31}}, \quad 
\frac{F_v}{v} \sim \left(\frac{\Lambda^{84}}{M_P^{53}} 
\right)^{\frac{1}{31}}. 
\end{equation}
Since the vacuum energy is non-zero, 
SUSY is certainly broken. 
This breaking effect 
is communicated to the observable sector 
as follows. 
Upon identifying the $SU(5)$ subgroup of 
the flavor group $SU(6)$ with our usual 
gauge group which includes 
the SM gauge groups, 
$Q$ and $\bar{Q}$ behave as 
${\bf 5}+{\bf \bar{5}}$ 
messenger fields of which SUSY mass is 
$\lambda_1 v$ and SUSY breaking (mass$)^2$ 
is $\lambda_1 F_v$. 
They communicate SUSY breaking to the soft 
masses through loop diagrams of $Q, \bar{Q}$ 
in the usual way \cite{DN}. 
Note also that since the original model 
is completely chiral, 
these vectorlike messengers are dynamically 
generated 
as a result of symmetry breaking. 
Furthermore, there is no gauge messengers 
and no light scalars charged under the SM 
which gives a negative contribution to the 
soft (mass$)^2$ since $\Sigma$ is a singlet 
for $SU(5)$.

Requiring $F_v/v\sim 10^4$ GeV to obtain 
the soft masses of order $10^{2\sim3}$ GeV, 
we find 
\begin{eqnarray}
\Lambda &\sim& 7 \times 10^{12} {\rm GeV}, 
\quad 
\Lambda_L \sim 1 \times 10^{12} {\rm GeV}, 
\nonumber \\ 
v &\sim& 3 \times 10^{14} {\rm GeV}, 
\quad 
\sqrt{F_v} \sim 1 \times 10^9 {\rm GeV}, 
\end{eqnarray}
where we used $M_P = 2 \times 10^{18}$ GeV.

We give some comments on the above scales 
in order. 
Firstly, since $F_v < v^2$, the SM gauge groups 
are not broken at the minimum. 
Secondly, the messenger scale $v$ is close to 
the GUT scale, so it is possible 
to preserve the perturbative unification 
in spite of six flavors of messengers. 
Thirdly, one may worry about 
that the SUGRA contribution 
to the soft masses is comparable or dominant 
because SUSY breaking scale $\sqrt{F_v}$ is 
relatively large. 
If we require that the gravitino mass $m_{3/2}$ 
which is the typical scale of the SUGRA contribution 
is less than 10 percent of the gluino mass, 
\begin{equation}
m_{3/2} = \frac{F_v}{\sqrt{3}M_P} < 
0.1 \times 6 \times \frac{\alpha_s}{4\pi} 
\frac{F_v}{v}, 
\end{equation}
then we find 
\begin{equation}
v < 0.1 \times 6 \times \frac{\alpha_s}{4\pi} 
\sqrt{3} M_P \sim 2 \times 10^{16} {\rm GeV}. 
\end{equation}
In our model, the above requirement is clearly 
satisfied, so the SUGRA contribution is 
suppressed enough.

Recall here that there remain still classical 
flat directions $B, \bar{B}$ and $M$. 
We have to argue whether or not these directions 
are lifted quantum mechanically.

Along the $B$ direction, 
$SU(6)_1$ and $SU(6)$ are completely broken. 
Then $\Sigma$ and $\bar{Q}$ become massive, 
hence the low energy effective theory is 
$SU(6)_2 + {\rm singlets}$. 
The dynamical superpotential is 
\begin{equation}
W_{dyn} = \Lambda_L^3 = \Lambda^2_2 B^{1/6}, 
\end{equation}
where we use the scale matching 
$\Lambda_L^3 = \Lambda_2^2 B^{1/6}$. 
This leads to non-zero constant vacuun energy. 
At one loop, the correction to K\"ahler potential 
makes the scalar potential stabilized 
near the origin \cite{Murayama,DDRG,Shirman2}.

Along the $\bar{B}$ direction, 
$SU(6)_2$ and $SU(6)$ are completely broken. 
Then $\Sigma$ and $Q$ become massive, 
hence the low energy effective theory is 
$SU(6)_1 + \antithree +$ a singlet. 
The effective superpotential becomes
\footnote{We also assume here that the theory is 
in the $W_{dyn}=0$ branch.} 
\begin{equation}
W_{eff} = \frac{\lambda_2}{M_P} u
\end{equation}
Using the canonical K\"ahler potential 
for $u$ and the scale matching, 
we obtain the scalar potential 
\begin{equation}
V_{eff} = \frac{\Lambda_L^6}{M_P^2} = 
\frac{\lambda_2^2}{M_P^2} \Lambda_1^{18/5} 
\bar{B}^{2/5}. 
\end{equation}
Clearly, this stabilizes $\bar{B}$ direction.

Along the $M$ direction, $SU(6)_1$ and $SU(6)$ 
are completely broken. 
The low energy effective theory is 
$SU(6)_2 + {\rm singlets}$. 
This is the same effective theory 
along the $B$ direction. 
Therefore, the $M$ direction is also lifted.

We now briefly dicuss the drawback of our model. 
In the case with $W_{dyn} \ne 0$, 
this $W_{dyn}$ becomes a runaway potential 
for $u, v$. 
Then one can easily see that SUSY vacuum exists. 
Therefore, our model does not work in this case 
and it is inevitable to assume that 
$SU(6) + \antithree$ model is in $W_{dyn}=0$ 
branch. 
The same assumption is needed 
if we apply the model with $\mu < \mu_{adj}$ 
($\mu, \mu_{adj}$: Dynkin index of matter 
representation, that of the adjoint 
representation.) in Ref. \cite{DM} 
instead of 
$SU(6) + \antithree$ model. 
On the other hand, 
if we apply the model with $\mu > \mu_{adj}$ 
in Ref. \cite{DM}, 
we do not have to assume $W_{dyn}=0$ 
because $W_{dyn} \ne 0$ branch does not exist 
in this case\footnote{Shirman \cite{Shirman} 
uses the model that has no flat directions 
as SUSY breaking sector in the first example. 
There is also no need to assume $W_{dyn}=0$ 
because it is trivial in this case.}. 
For instance, it is an interesting 
challenge to construct a model with 
direct gauge mediation by using ISS model 
\cite{ISS} as SUSY breaking sector.

In summary, we have presented a new, simple 
SUSY breaking model with direct gauge mediation. 
We have utilized the model with an Affine quantum 
moduli space as SUSY breaking sector. 
The model is completely chiral 
and has no gauge messengers and 
no light scalars charged under the SM gauge 
groups which gives a negative contribution 
to the soft (masses$)^2$. 
Since messenger fields are so heavy, 
the perturbative unification can be preserved. 
We have estimated the characteristic scales 
and obtain phenomenologically desirable values. 
Furthermore, we have shown that 
the SUGRA contribution is suppressed enough.

\begin{center}
{\bf Acknowledgements}
\end{center}
The author would like to thank C. Cs\'aki 
for valuable comments. 
The author also would like to thank S. Kitakado 
for careful reading of the manuscript. 
This work was supported by Research Fellowships 
of the Japan Society for the Promotion of Science 
for Young Scientists.

%\newpage

\end{document}